\documentstyle[aps,pre,eqsecnum,preprint]{revtex}

\begin{document}
\bibliographystyle{unsrt}
\textwidth 800pt
\large
\begin{center}
\underline{Interaction of discrete breathers with electrons
in nonlinear lattices}  
\vspace{2cm}\\ \large S. \vspace{0.5cm}Flach$^*$ and K. Kladko \\
\normalsize
Max-Planck-Institut f\"ur Physik komplexer Systeme \\
Bayreuther Str.40 H.16, D-01187 Dresden, Germany \vspace{1cm} 
\\
\end{center}
\normalsize
ABSTRACT \\
{ \small We study the effects of electron-lattice interaction 
in the presence of discrete breathers.
The lattice is treated classically. 
We consider two different situations - i) the scattering of
an electron by a discrete breather in the 
semiconducting regime,
where the electron-breather distance is large compared to
the breather size, and ii)
the appearance of a bound electron-breather state, which exists
at least over one half of the breather period of oscillation.
In the second case the localization length of the electron
can be of the order of the breather size - a few lattice periods.
Remarkably these results are derived in the absence of disorder, since
discrete breathers exist in translationally invariant nonlinear lattices,
}
\vspace{0.5cm}
\newline
PACS number(s): 03.20.+i ; 03.65.Ge; 03.65.Nk; 63.20.Kr; 63.20.Ry; 72.10.Di 
\newline
\newline
\\
\\
\\
$^*$ email: flach@idefix.mpipks-dresden.mpg.de
\newpage

\section{Introduction}
 
The concept of discrete breathers has been studied in detail in
a number of publications
\cite{st92},\cite{jbp90},\cite{cp90},\cite{fw3},\cite{fw8},\cite{fw7},\cite{fw9},\cite{ma94}. 
Discrete breathers are time-periodic solutions
of a system of coupled classical degrees of freedom, typically arranged on
a translationally invariant lattice.
These solutions are spatially localized (note that spatial localization
appears in the absence of any disorder). 
 The number of degrees of
freedom can be finite or infinite. A necessary condition for the appearance
of discrete breathers is the presence of nonlinear terms in the
equations of motion. A discrete breather can be viewed as a localized
excitation of the system above its classical ground state. Localization
occurs because multiples of the frequency of the breather can easily escape
from resonances with the spectrum of the linearized equations of
motion (around the ground state). This happens because the linear spectrum
of the system (e.g. the phonon spectrum of a crystal) has a finite upper
bound due to the discrete translational invariance (as opposed to
a system with a continuous translational invariance, e.g. a field
equation). 
Note that no specific topological requirements have to be met
in order to obtain breather solutions, especially there are no restrictions
with respect to the dimension of the lattice.

Although the concept of discrete breathers goes much beyond
the description of nonlinear lattice dynamics in crystals, we will
focus our attention in this work 
to crystals only. In particular we will study the
problem of interaction of a discrete breather with electronic degrees
of freedom. We will consider systems where screening effects due
to mobile electrons are weak. In section II we derive the scattering
of an electron by a discrete breather in the limit of large 
electron-breather distance. In section III the trapping of an electron
by a discrete breather is analyzed. Remarkably the corresponding
localization length of the electron can be of the order of the 
breather size, i.e. a few lattice sites.

\section{Scattering of an electron by a discrete breather}

Let us consider a classical nonlinear lattice which allows for
breather solutions. 
Generally the excitation of the relevant breather degrees of freedom
leads to a localized polarization of the lattice.
In the classical ground state of this
system 
the polarization vanishes.
If we excite a discrete breather, then it will induce
a (time-periodic) multipole field at distances large compared
to the breather size. Generally 
the first nonvanishing momentum will be a
dipole momentum. The induced polarization will be spatially
localized, in accord with the strong localization properties of
the breather solution.

Let us consider the interaction of a single electron with a discrete
breather in the case when the distance between the electron and the breather
is much larger than the breather size. 
Since we are describing the lattice degrees of freedom classically,
we can use the adiabatic approximation \cite{asd73}. This means, that
the motion of the electron is described by using the positions
of the lattice degrees of freedom as parameters. Thus the electron feels
a multipole field which originates from the breather. This kind of treatment
of the electron is similar to the consideration of electrons in a lattice
with defects \cite{jm73}. 
The difference is, that i) the breather (dynamical defect)
does not posess a uncompensated charge and ii) the breather is
slowly (as compared to the inverse electron frequency) changing its multipole
field. 

Since the multipole field of the breather contains in general a dipole
component, we can study the scattering of an electron in a 
dipole field. We consider the case when  the 
electron can follow a path which does not
come close to the breather location. If this assumption is not
true anymore, the electron can be trapped by the breather, as will be
shown in the next section.

The potential of a dipole is given by
\begin{equation}
V_d(\vec{r}) = \frac{1}{\epsilon}\frac{\vec{d}\vec{r}}{r^3}\;\;.\label{2-1}     
\end{equation}
Here $\vec{d}$ is the dipole moment induced by the breather (which
is actually slowly periodically oscillating with time). The dielectric
permeability $\epsilon$ describes the reduction of the dipole
field due to polarization effects.

The motion of an electron with isotropic effective mass
$m^*$ and charge $e$ will be then 
described by the Hamiltonian $H$ and the wave function $\Psi(\vec{r},t)$
\cite{jm73}
\begin{equation}
H = -\frac{\hbar ^2}{2m^*} {\rm \Delta} + eV_d(\vec{r}) \;\;,
\;\; i\hbar \frac{\partial \Psi}{\partial t} = H \Psi\;\;.
\label{2-2}       
\end{equation}
The dipole potential (\ref{2-1}) does not posess localized states.
This can be easily found by considering the corresponding classical
motion in the potential (\ref{2-1}). Clearly there exist no periodic
orbits having some finite distance from the potential center $\vec{r}=0$.
Thus there appear no localized
electron states which are weakly localized as compared to the breather
size.This situation is opposite to the Coulomb field problem where periodic
orbits exist and lead to the existence of hydrogen like localized states,
as used in the description of Wannier excitons. To find trapped electronic
states induced by a discrete breather we have to take into account
the internal breather structure, which will be studied in the next section.

To account for the elastic electron reflection in the 
dipole potential (\ref{2-1})
we can use Born's approximation \cite{asd73}, 
which holds if the interaction energy
between the electron and the breather will be small compared to the
kinetic energy of the electron. Denoting by $|\vec{k}>$ the plane wave
states of the electron in the absence of a breather, we have to 
calculate the matrix elements
\begin{equation}
<\vec{k}|V_d|\vec{k}'> = \int {\rm e}^{i(\vec{k}-\vec{k}')\vec{r}}
V_d(\vec{r}){\rm d}r^3\;\;.\label{2-3}         
\end{equation}
Straight integration gives
\begin{equation}
<\vec{k}|V_d|\vec{k}'> = -i \frac{4e \pi}{\epsilon} 
\frac{\vec{d}(\vec{k}-\vec{k}')}
{|\vec{k}-\vec{k}'|^2} \delta(E_{\vec{k}} - E_{\vec{k}'})\;\;. \label{2-4}      
\end{equation}
The electronic energies $E_{\vec{k}}$ measure the energy of the incoming
and outgoing plane waves.                    
All other quantities related to the electron scattering in the used
approximation can be obtained from these matrix elements.

In the nongeneric symmetric case that the breather 
does not posess a dipole momentum,
the quadrupole field tensor $D_{\alpha \beta}$ has to be considered
(note that we use the definition $D_{\alpha \beta}=\sum_i e_i x^{(i)}_{\alpha}
x^{(i)}_{\beta}$).
The corresponding potential is given by
\begin{equation}
V_q(\vec{r}) = \frac{1}{2\epsilon} D_{\alpha \beta} \frac{\partial}
{\partial x_{\alpha}}\frac{\partial}{\partial x_{\beta}} \frac{1}{r}
\;\;. \label{2-5}       
\end{equation}
Again there are no bound states in potential (\ref{2-5}) as in the dipole
case. The matrix elements can be obtained by integrating:
\begin{equation}
<\vec{k}|V_q|\vec{k}'> = -i \frac{2e \pi}{\epsilon}
 \frac{D_{\alpha \beta}k_{\alpha}k_{\beta}}{|\vec{k}-\vec{k}'|^2} 
\delta(E_{\vec{k}} - E_{\vec{k}'})\;\;. 
\end{equation}

Let us comment on possible numerical and experimental
verifications of the results from this section. Numerical
investigations can be performed if the motion of the electron
is described within a tight-binding representation (cf. next
section). For one-dimensional systems the task amounts
to launching an electron wave with given wave number and accounting
for its collision with a breather. The dependence of the transmission
coefficient on the wave number can be then compared with our results.
Similar studies were done for the scattering of phonons by breathers
\cite{fwdna}. A numerical analysis of the two-dimensional case is also possible
along the same lines, although it will be much harder and tedious.
As for experiments, up to now there is no substantial knowledge
on how to excite breathers, and how to provide a statistical description
of scattering processes in the presence of breathers. Once the new 
collision integral \cite{aaa88} is obtained, one can think of possible
experimental realizations to check the influence of breathers
on the electrical conductivity.

\section{Trapping of an electron by a discrete breather}

In this section we analyze the properties of localized electronic states,
induced by the presence of a discrete breather. We restrict the consideration
to the simplest case of a one-dimensional tight-binding description of the
electron \cite{ene83}, 
where the overlap integral of the electron states at neighbouring
sites depends on the positions of the corresponding crystal atoms.

Again we describe the crystal atoms classically, so that we have to
use the adiabatic approximation.
Consequently we can treat only 
electronic states, which are energetically well separated from other
electronic states. An electron sitting in such an adiabatic state will
continue to stay in this state if the Hamiltonian of the electron is changed
adiabatically slowly (due to the motion of the crystal atoms). Clearly
we can not consider extended electronic states, which usually form
a band (so that the energy separation is violated). The idea is then
to show that at some intervals in time the localized lattice distortion due
to the presence of a discrete breather allows for localized electronic states,
which are well separated from the electronic band. If an electron is
occupying such a localized state, the dynamics of the breather will
be changed according to the adiabatic approximation.

The Hamiltonian $H$ of the system is given by
\begin{eqnarray}
H_{lat}= \sum_{l} \left[ \frac{1}{2} P_l^2 + V(X_l) + \Phi(X_l - X_{l-1})
\right] \;\;, \label{3-1} \\
H_{el} = \sum_l \beta_{l,l+1} (a^+_la_{l+1}+ cc)\;\;,\;\;
\beta_{l,l+1}=1 + \beta_1
(X_l - X_{l-1})\;\;, \label{3-2} \\
H = H_{lat} + H_{el}\;\;. \label{3-3}       
\end{eqnarray}
The lattice part $H_{lat}$ contains anharmonic terms $(\partial^2 H_{lat}
/ \partial X_l \partial X_{l'} \neq {\rm const})$ in the potential functions
$V(z)$ and $\Phi(z)$. The electron-lattice coupling is considered in linear
approximation in the overlap integral $\beta$. The one-site electronic
energies are assumed to be independent on the lattice site number $l$,
so that the corresponding terms $a^+_la_l$ can be scaled away in (\ref{3-2}). 

The adiabatic approximation implies to solve the eigenvalue problem
for the electron (\ref{3-2}) using the lattice degrees of freedom as 
fixed parameters. 
The electronic band for (\ref{3-2}) is given by
\begin{equation}
E_k = 2 {\rm cos}(k) \label{3-4}
\end{equation}
(here $k$ measures the wavenumber of the extended electron wave). 

To account for the localized state of the electron in the presence
of a discrete breather we arrive at the equations for the probability
amplitudes $c_l$ of finding the electron at lattice site $l$:
\begin{equation}
E_{el}c_l = \beta_{(l-1),l}c_{l-1}+\beta_{l,(l+1)}c_{l+1}\;\;. \label{3-5}       
\end{equation}
In order to proceed we have to use a known discrete breather solution.
It is known that breather solutions
exist, when essentially one (central) atom is performing periodic 
oscillations, with all the other atoms having exponentially small amplitudes
\cite{fw7},\cite{ma94},\cite{fw10}.
Discrete breathers with such strong
localization properties were found in numerical studies 
for one-dimensional and two-dimensional
lattices with moderate coupling and anharmonicity \cite{fw3},\cite{fw8}. 
Thus we assume the
easiest form of the breather solution to be, that only one central particle
is oscillating (with a frequency outside the linear phonon spectrum).
Taking into account the small (compared to the central atom) amplitudes
of vibration of the rest of the lattice leads to small corrections of
the results derived below.

If the breather is essentially given by the motion of one atom $X_0(t)=
X_0(t+T_b)$,
then we can solve analytically for the electronic state. All other atoms
are at their groundstate positions $X_{l \neq 0}=0$. Solving (\ref{3-5})
to the left and right of the breather location yields
\begin{eqnarray}
c_l = c_1 {\rm e}^{-\lambda l}\;\;, \;\;l>1\label{3-6} \\
c_l = c_{-1}{\rm e}^{\lambda l}\;\;,\;\;l < -1 \label{3-7} \\
E_{el}= 2{\rm cosh}(\lambda)\;\;. \label{3-8}           
\end{eqnarray}
To obtain the energy $E_{el}$ we have to match the two branches of
our solution in the breather center, when the amplitude of the central
atom at a given time is $X_0(t)$. The result is
\begin{equation}
E_{el} = \frac{1}{\sqrt{1 + 2\beta_1^2X_0^2(t)}} + \sqrt{1 + 
2\beta_1^2X_0^2(t)}
\approx 2 + \beta_1^4X_0^4(t) + O(\beta_1^6X_0^6(t))\;\;.\label{3-9}      
\end{equation}
The exponent of the spatial decay of the electronic localized state
$\lambda$ is then given by
\begin{equation}
\lambda = \frac{1}{2}{\rm ln}(1 + 2\beta_1^2X_0^2(t))\;\;.\label{3-10}       
\end{equation}
The unnormalized probability amplitudes in the center of the breather
location are given by
\begin{eqnarray}
c_0 = 1\;\;,\label{3-10a} \\
c_1=\sqrt{1 + 2\beta_1^2X_0^2(t)}(1-\beta_1X_0(t))\;\;,\label{3-10b} \\
c_{-1}=\sqrt{1 + 2\beta_1^2X_0^2(t)}(1+\beta_1X_0(t))\;\;,\label{3-10c}       
\end{eqnarray}
The dynamics of the discrete breather is now modified due to the
additional purely anharmonic 
electron-induced potential (\ref{3-9}). Depending on
the breather solution (frequency above or below the phonon band) this
additional potential can either support the localized breather solution
or supress it \cite{fw10}. 
Remarkably there {\sl is} a realization when it can support
the breather solution - that means that the breather gets even more strongly
localized, inducing again a more strong localization of the electronic
state etc. 

It is worthwhile to mention a numerical study, where the lattice itself
does not support discrete breathers, but a bound electron-breather
state exists \cite{vr94},\cite{vr95}. We can give a simple explanation
to this observation. 
The relevant expansion parameters 
for the potential
$\Phi(z)=\sum_{\mu=2}^{\infty}\frac{1}{\mu}
\phi_{\mu}(z-z_0)^{\mu}$ in \cite{vr94},\cite{vr95} 
are given by $\phi_2=0.005625$, $\phi_3=
-0.004922$, $\phi_4=0.001812$.
It has been shown recently, that breather solutions do not
exist for that lattice, because the inequality $3/4\phi_2\phi_4 \geq \phi_3^2$
is not fulfilled. 
This steems from the fact, that the corresponding upper zone boundary
plane wave undergoes a tangential bifurcation only if the above inequality
{\sl is} fulfilled \cite{fw11},\cite{sp94}.
However the correction of the lattice potential energy
due to a (initially) localized electron can change this relation locally
(cf. (\ref{3-9})),
so that breathers can exist if a localized electronic state is occupied.

Since $X_0(t)$ of the breather solution becomes periodically zero,
the adiabatic approximation will become invalid every half-period of
the breather's oscillation. 
That happens because the localized state of the electron merges with
the electron band states.
So we conjecture  that a trapped (localized)
electronic state may exist only over roughly every half-period of the breather.
That observation follows from the circumstance that the adiabatic
approximation holds if the separation of the localized state from the
continuum is much larger than the change of interaction energy during
one period of the electronic amplitude oscillations which implies
\[
\beta_1^4X_0^5(t) \gg \dot{X}_0(t)\;\;.
\]
Each time when the central atom will pass its stability position $X_0=0$
the localized electronic state will come close to the electronic band.
Consequently some part of the initially localized electron wave
function will be emitted into the electronic band. This will lead to
a finite life time of the localized electronic state of the order of
one half of the breather period (this can be also seen in the numerical
result in \cite{vr94},\cite{vr95}).

We also briefly mention the case of another  breather symmetry, 
when a breather is given
essentially by the motion of two nearest neighbours $X_{0}(t)=-X_1(t)$.
These two-site breathers appear with frequencies above the linear spectrum
(phonon band). Relations (\ref{3-6})-(\ref{3-8}) hold again.  
The energy of the localized electronic state is given by
\begin{eqnarray}
E_{el} = F + \frac{1}{F}\;\;,\label{3-11} \\
F = \frac{2x+1-\sqrt{1+8x^2-4x}}{2x(2-x)}\;\;,\;\;
x=\beta_1X_0(t)\;\;.\label{3-12}  
\end{eqnarray}
Note that in lowest order of $\beta_1X_0(t)$ the energy again becomes
\begin{equation}
E_{el} \approx 2 + \beta_1^4X_0^4(t) + O(\beta_1^5X_0^5(t))\;\;. \label{3-13}       
\end{equation}

Here we have considered highly discrete (localized) breather solutions
only. Taking into account the (nonzero) oscillations of the other
atoms in the breather solution will smoothly modify our findings, but
can not change them qualitatively (unless the breather becomes very
weakly localized). 

Let us consider the case of higher dimensional lattices. Discrete breathers
can be obtained in full analogy to the one-dimensional case. However
the localization properties of the electron will be modifed. In two-dimensional
lattices the separation of the localized electronic state from the
electronic band will be exponentially small for small breather amplitudes
\cite{LLIII}.
In three-dimensional lattices the discrete breather amplitude has even
to exceed a certain finite threshold in order to localize an electron
\cite{LLIII}.
So in higher dimensional lattices the trapping time of an electron will
be shortened as compared to the one-dimensional case.

\end{document}